# Thermal conductivity of $CdCr_2Se_4$ ferromagnet at low temperatures: role of grain boundaries and porosity

Jiří Hejtmánek[1*], Kyo-Hoon Ahn[1], Zdeněk Jirák[1], Petr Levinský[1], Jiří Navrátil[1], Sandy Al Bacha[2], Emmanuel Guilmeau[2] and Karel Knížek[1]

[1]Institute of Physics of the Czech Academy of Sciences, 162 00 Prague 6, Czech Republic.
*Corresponding author:* hejtman@fzu.cz

[2]CRISMAT, CNRS, Normandie Univ, ENSICAEN, UNICAEN, 14000 Caen, France

**ABSTRACT:**

It is unambiguously demonstrated that the low temperature magnon specific heat in a ferromagnet varies as $T^{3/2}$ and the magnon thermal conductivity, due to $T^{1/2}$-dependent effective velocity of magnons, as $T^2$. The confirmation of these model comportments is based on the experimental study of chalcospinel $CdCr_2Se_4$, which represents relatively rare example of a ferromagnetic insulator ($T_C$=130 K) without undesirable "masking" contributions of the itinerant electron excitations and nuclear specific heat that both make impossible to conclusively unveil the role of magnons. The ratio of the magnon to lattice specific heat is found to reach 87:13 at 2 K and is in accordance with predictions based on the spin-wave stiffness $D$ = 33.5 meVÅ$^2$ and Debye temperature $\theta_D$ = 237 K. On the other hand, the ratio of the magnon to phonon thermal conductivity reaching 27:73 at 2 K is much lower than expected for standard model of the grain boundary limited transport. This suggests that mean free paths for long-wavelength magnon/phonon heat carriers are largely different – shorter than the grain size (of ~1μm) for magnons and longer than grain size for phonons. The phonon dominated low temperature thermal conductivity exhibits, moreover, a ~$T^{2.3}$ temperature dependence instead of the standard predicted model in $T^3$. The relevant scattering mechanisms, both the phonon frequency-independent and -dependent ones, are discussed in detail.

## INTRODUCTION

Although the low-temperature thermal properties of ferromagnets are routinely interpreted using textbook theoretical models, the contribution of spin-wave excitations (magnons) is difficult to be clearly assessed and analyzed experimentally, namely due to more significant contribution of lattice excitations (phonons). Phonon and magnon contributions become clearly distinguishable only at very low temperatures where, unfortunately, they are often masked by other pertinent contributions: namely that of electronic origin (itinerant electrons in metallic systems), (ii) Schottky type (both electronic and nuclear) or (iii) "extrinsic-like" associated with phase purity or sample morphology (e.g. glassy state artificially boosting the low temperature heat capacity). In this report we have chosen as a model compound the insulating ferromagnetic selenospinel $CdCr_2Se_4$ ($T_C$~130 K), in which all above-mentioned inconveniences are eliminated.

The $CdCr_2Se_4$ phase is characterized by spinel structure of cubic $Fd\bar{3}m$ symmetry with the non-magnetic $Cd^{2+}$ cations in tetrahedral sites and magnetic $Cr^{3+}$ cations in octahedral sites. In such a case, there are generally two competing interactions between nearest-neighbor transition cations, the antiferromagnetic direct 3d-3d exchange and the 90° ferromagnetic superexchange via anionic p-orbitals, each of specific dependence on orbital overlaps [1,2,3]. While closely related oxospinels $ZnCr_2O_4$ or $CdCr_2O_4$ are frustrated antiferromagnets, the present system possessing large $Se^{2-}$ ions and reduced direct exchange is a stable ferromagnetic insulator [4,5]. Concerning the thermal properties, it should be stressed that here selected compound does not contain any element possessing noticeable low temperature nuclear $C_p$ contribution nor electronic Schottky contribution (the $t_{2g}^3$ orbital singlet state of $Cr^{3+}$ yields an ideal spin moment 3 $\mu_B$/Cr lacks any close-energy excited states). Finally, the risk of impurities/glassy phase formation is minimized as phase diagram enables the synthesis of pure phase with minimum contamination by impurities. Moreover, the cubic symmetry, moderate Curie temperature and absence of elements with significant spin-orbit coupling anticipate the possibility to suppress the magnons by external magnetic field. We note that, in the case of other well-known FM insulators, like EuO or other based on rare-earth metals, the noticeable nuclear contribution to the specific heat below 1 K makes the analysis of magnons difficult [6].

As the $CdCr_2Se_4$ system and its electronic and magnetic properties are concerned, the detailed theoretical survey, *ab-initio* calculations and basic experimental characterization have been presented in our previous publication [7]. Here we focus on the low-temperature specific heat capacity and thermal conductivity of ceramic $CdCr_2Se_4$ samples and exemplify the suppression of respective magnon-controlled contributions by application of a strong magnetic field. Due to insulating character of $CdCr_2Se_4$, the specific heat capacity is composed of only two terms – the contribution of transversal and longitudinal acoustic lattice modes, and the contribution of single acoustic mode of spin-wave excitations. The same bosonic particles (phonons and magnons) contribute as heat carriers to the thermal conductivity. We examine especially the effects of grain boundaries and sample porosity that are decisive for thermal transport of polycrystalline $CdCr_2Se_4$ below 10 K.

**EXPERIMETAL**

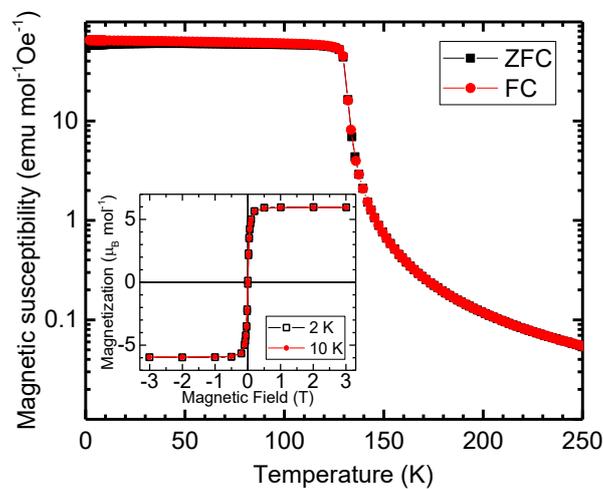

Fig. 1. The temperature dependence of magnetic susceptibility of $CdCr_2Se_4$ measured at DC field of 100 Oe. In the inset, the magnetization reaches 6 $\mu_B$/f.u. and confirms ideal ferromagnetic state at low temperatures.

Ceramic samples CdCr$_2$Se$_4$ were prepared from a stoichiometric mixture of intimate mixture of pure elements. The mixture was heated in evacuated silica ampoule up to 1073 K, then quenched into cold water, re-ground and heated again in an evacuated ampoule up to 973 K for another 5 days. The final powdered product was sintered using either

(i) Hot-pressing at 763 K and pressure of 50 MPa for 1 hour which resulted in relatively porous ceramics with relative density 83%. Sample is further labelled as porous.

(ii) Spark plasma sintering (SPS) where the powder was placed into tungsten carbide dies and densified using SPS (SPS-FCT HPD 25) at 773 K for 30 min under a 300 MPa pressure (24 KN). The produced pellets were of a 10 mm diameter and a thickness of ~ 6.5 mm with geometric densities exceeding 95 %. Considering extreme sintering conditions, the sintering was followed by annealing at 500 °C in selenium atmosphere for 24 hours. The sample is further labelled as dense.

The annealing in selenium atmosphere enabled to recover the high phase purity of the ceramics, similar to hot-pressed sample i.e. with main impurities identified both by energy-dispersive X-ray spectroscopy (EDS) and X-ray diffraction (XRD) as CdSe and Cr$_2$Se$_3$ with concentrations below 1% of both- see Supplementary information. The phase purity of both samples was checked by powder X-ray diffraction acquired on Bruker D8 Advance powder diffractometer with CuK$\alpha$ radiation equipped with Lynxeye XE-T detector. Microstructure of the samples was analysed by scanning electron microscopy and microanalyzer Jeol JXA -8230 with EDS Bruker QUANTAX 200 (software Esprit 2.5 ED). The average grain size larger than 1 µm deduced from the width of diffraction peaks agrees with the SEM images showing grains with sizes of the units of µm (see the Supplementary information). The EDS analysis of the dense ceramics revealed average composition of Cd$_{1.01(0.02)}$Cr$_{2.02(0.06)}$Se$_{3.97(0.04)}$ which is very close to ideal spinel stoichiometry – see Table S1.

The magnetic response was measured using a superconducting quantum interference device (SQUID) magnetometer (MPMS-XL, Quantum Design), the thermal properties were characterized by means of the Physical Property Measurement System (PPMS, Quantum Design). For more detailed description, please, see Ref. 7.

**RESULTS and DISCUSSION**

**Fundamentals of polycrystalline CdCr$_2$Se$_4$; magnetism and thermal properties**

Ferromagnetic ordering of the studied CdCr$_2$Se$_4$ is demonstrated in the Fig. 1. The negligible difference between the FC (field cooled) and ZFC (zero field cooled) magnetic susceptibility confirms the good magnetic homogeneity without any tendency to form clusters or inhomogeneities. The ideal ferromagnetic state is corroborated by the magnetic moment corresponding to 3 µ$_B$/Cr, which value perfectly matches the spin-only contribution of octahedrally coordinated Cr$^{3+}$ ions.

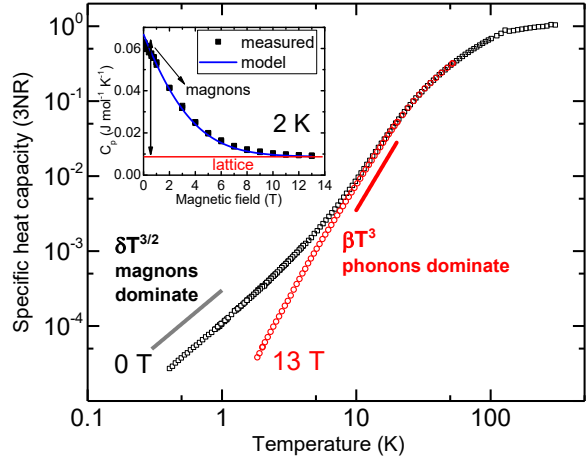

Fig. 2. The temperature dependence of CdCr$_2$Se$_4$ specific heat capacity expressed as relative value with respect to the Dulong-Petit limit. The data at zero field (black) and $\mu_0 H$=13 T (red) are shown. In the inset, the field induced suppression of magnon contribution to the specific heat is evidenced at 2 K where magnetic contribution represents 87% of total heat capacity.

The temperature dependence of the specific heat capacity is shown in Fig. 2. The ferromagnetic transition is characterized by a small peak at 130 K; the faint feature detected at $T_C$ corroborates its 2-nd order character. The logarithmic scale of Fig. 2 enables us to visualize the low-temperature behavior, where the comparison between the data measured at 0 T and 13 T clearly uncovers the huge impact of the magnetic field on the heat capacity. This is further highlighted in the inset, where the magnetic field dependence of the heat capacity at 2 K shows a gradual suppression of the specific heat and its final saturation under 13 T at only ~13% of its original value. Consequently, the specific heat measured within the range 0.35-5 K displays under zero magnetic field the temperature dependence $C_p \sim T^{3/2}$, which perfectly agrees with theoretical predictions for a simple cubic ferromagnet magnon-specific heat, and changes under strong magnetic field to the classic phonon dominated dependence $C_p \sim T^3$ (see the main panel of Fig. 2). To better visualize the relative impact of lattice and spin waves at low temperatures, we depict the low temperature specific heat capacity in a standard way as $C_p/T$ $vs.$ $T^2$ in Fig. 3.

Low temperature behavior of CdCr$_2$Se$_4$ agrees with the quantification of the specific heat capacity in cubic ferromagnetic insulators as a sum of the lattice and magnon terms. Molar lattice heat is given by Debye formula

$$C_{ph}(T) = \frac{12\pi^4}{5} n N_A k_B \left(\frac{T}{\Theta_D}\right)^3, \quad (1)$$

where $n$ is the number of atoms in formula unit ($n$ = 7 for CdCr$_2$Se$_4$), $N_A$ Avogadro number and $\theta_D$ Debye temperature. The molar magnon heat is derived for the case of a simple, gapless quadratic dispersion for spin waves, $E = \hbar\omega_k = Dk^2$, and is applicable at zero magnetic field only,

$$C_m(T) = 0.113 V_{mol} k_B \left(\frac{k_B T}{D}\right)^{3/2} \quad (2)$$

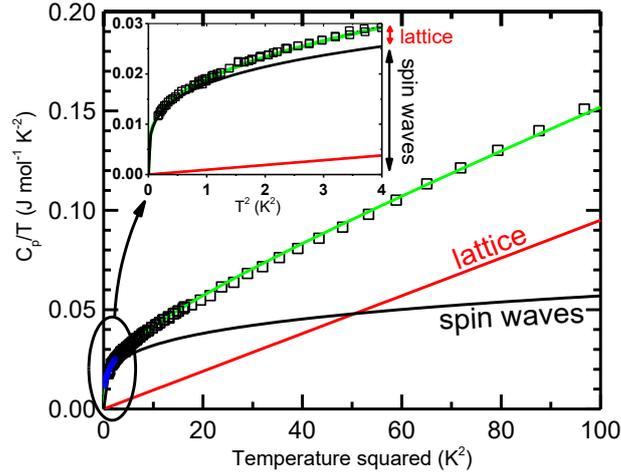

Fig. 3. The low-temperature specific heat capacity of CdCr$_2$Se$_4$ expressed in a standard way. The splitting between the lattice (red line) and magnon (black line) contributions provides a perfect agreement between theoretical expectations (green line) and experiment (open squares). In the inset, the experimental data down to 350 mK confirm the dominating contribution of magnons to specific heat capacity below 2 K; missing low temperature rise of specific heat corroborates the absence of the nuclear magnetic moments of stable isotopes of Cd, Cr and Se.

where $V_{mol}$ is molar volume (72.4·10$^{-6}$ m$^3$ for CdCr$_2$Se$_4$) and $D$ spin-wave stiffness, see e.g. Refs. 8-10. Under magnetic field, the magnons are gradually depopulated because of Zeeman gap $\triangle = g_J \mu_B \mu_o H$ for magnon excitations. Here, $H$ is the applied field corrected for demagnetization fields (for details see the macroscopic magnon theory in p.63 of Ref. 8), $\mu_0$ the permeability of vacuum, $\mu_B$ is Bohr magneton and $g_J \sim 2$ for transition metal ions of frozen orbital momenta. The gap brings about a gradual suppression of magnon heat capacity that scales as $H/T$. The calculated "quenching" curve can be found elsewhere [7,9]; it matches rather well the experimental curve seen in the inset of Fig. 2.

Confirming that the low temperature specific heat of CdCr$_2$Se$_4$ perfectly fits the theoretical models, we evaluate $C_{tot}(T) = \beta T^3 + \delta T^{3/2}$ with phonon term determined by $\beta$ = 1.02×10$^{-3}$ J·K$^{-4}$·mol$^{-1}$ and magnon term given by $\delta$ = 19×10$^{-3}$ J·K$^{-2.5}$·mol$^{-1}$. The value of $\beta$ corresponds to the Debye temperature $\theta_D$ = 237 K, which falls between 180 K obtained in Ref. 11 and 280 K in Ref. 12. Considering the quadratic magnon dispersion, the value of $\delta$ corresponds to the spin-wave stiffness $D$ = 33.5 meVÅ$^2$, which is in a perfect agreement with the values obtained from the spin-wave resonance technique ($D$=31.3 meVÅ$^2$ [13]) and previous heat capacity measurement ($D$=33 meVÅ$^2$ [12]).

We are turning now to thermal conductivity data of CdCr$_2$Se$_4$ displayed as log-log graph in Fig. 4 (for normal scale see Fig. 5(a) below). At ~50 K, the thermal conductivity develops a broad maximum reminding typical phonon peak that arises in single crystal materials with increasing temperature as a compromise between enlarged phonon population and more frequent phonon-phonon scattering. The thermal conductivity maximum appearing at ~ $\theta_D/5$ is of only a moderate value, which points to the heat transport limitations due to polycrystalline character of the sample and its porosity. At higher temperatures (above ~$\Theta_D/2$), it is generally considered that the mean free path of heat carrying phonons is essentially influenced by dissipative phonon-phonon Umklapp processes. Based essentially on the simple model of acoustic phonons, the thermal conductivity is expected to vary as $T^{-1}$. In reality, one should consider different dispersion relations of the longitudinal/transversal phonons and the presence of optical phonons that have often only minor contribution to heat transport but provide significant scattering channels for the acoustic phonon transport, for more information see e. g. Refs. 14,15. The *ab-initio* theoretical calculations of intrinsic thermal conductivity have been recently performed for Si of diamond structure [16] and CrN of distorted rock-salt structure [17], both with 3 optical phonon branches, or for a more complex case of the defect-chalcopyrite $CdGa_2Se_4$ with 18 optical branches [18]. Remarkably, all these calculations show the high-temperature thermal conductivity to follow very closely the ideal $T^{-1}$ trend. As the present $CdCr_2Se_4$ system is concerned, its spinel structure gives rise to phonon spectra with 3 acoustic and 39 optical branches [7], making eventual *ab-initio* calculation of phonon-phonon processes extremely tedious. Moreover, the thermal conductivity in Fig. 4 follows in the paramagnetic state a different trend, namely the $T^{-1/2}$ dependence which we tentatively associate with local spin correlations and interconnected atomic displacements that act as phonon scatterers above Curie temperature ($T_C \sim 130K$) [19]. We note, however, that the presence of additional scattering, namely point defect-like, may cause also similar deviation from $T^{-1}$ to the $\sim T^{-1/2}$ temperature dependence as shown already by Klemens [20]. In the case of our sample, we anyway favor the "magnetic" scenario since the deviation from the standard Umklapp dominated $T^{-1}$ temperature dependence starts sharply at $T_C$ (see Fig. 4 and 5). We present some more discussion on this subject at the beginning of next chapter.

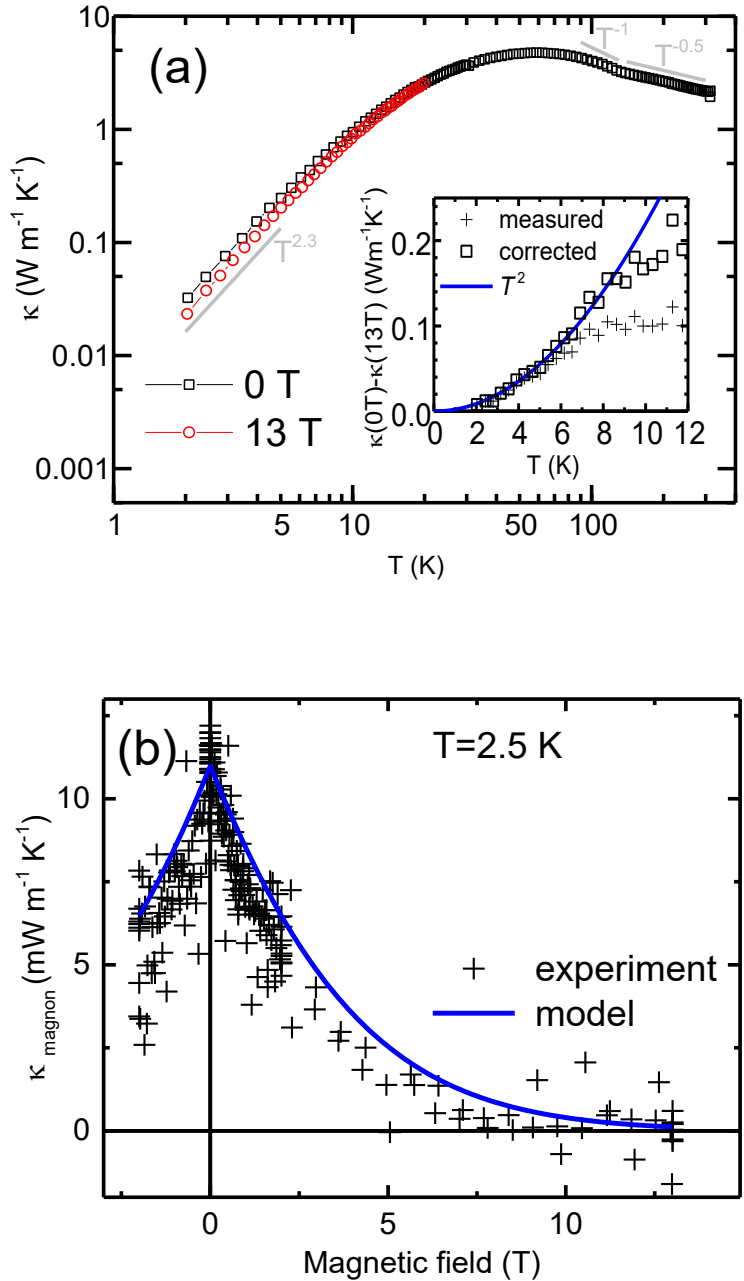

Fig. 4. The temperature dependence of the thermal conductivity of CdCr$_2$Se$_4$ (upper panel-a). The ceramic character of the sample (porosity 17%, grain size ~1 μm) is responsible for low temperature behavior which follows the ~$T^{2.3}$ dependence. In the inset we depict the magnon contribution to the thermal conductivity $\kappa_m$ ~ $T^2$. Magnon contribution $\kappa_m$ was determined as a difference between $\kappa_{0T}$ and $\kappa_{13T}$ (crosses) with additional correction for incomplete quenching of magnon contribution above 6 K (squares). The lower panel (b) shows the suppression of the magnon thermal conductivity at 2.5 K; complete quenching of magnon thermal conductivity at 13 T is supposed.

Our primary concern is, however, the behavior of thermal conductivity well below the phonon peak, especially with respect to the phonon/magnon contributions $\kappa_{ph}$ and $\kappa_m$. As seen in Fig. 4, the thermal conductivity follows the $T^{2.3}$ dependence and only a minor suppression is observed under field of 13 T. The isothermal field dependence, demonstrating the gradual quenching of $\kappa_m$, is shown for 2.5 K in the lower panel (for more data and discussion of $\kappa_m(H,T)$ see Ref.7). The difference between 0 and 13 T is presented in the inset in Fig. 4a. It should be noted that the quenching at 13 T becomes incomplete with increasing temperature (e. g., it amounts to 92% at 4K and 68% at 8 K, respectively) [7,10]. The data after proper correction follows rather well the $T^2$ dependence up to about 8 K. The deviation observed at higher temperatures will be commented in more detail at the end of next chapter; we just note that, instead of quasi-exponential quenching exemplified for 2.5 K and scaled as $H/T$, there is certain drop of thermal conductivity in low fields, followed with nearly constant value in 1 – 13 T.

In a first approximation, we confront the data with predictions based on the kinetic gas theory [8,19,21]. Neglecting the role of grain boundaries, such a conventional model gives the low-temperature thermal conductivity of ferromagnetic insulator as a sum of lattice and magnon terms, $\kappa = \frac{1}{3}C_{ph}v_{ph}l_{ph} + \frac{1}{3}C_m v_m l_m$. Here, $C_{ph}$ ($C_m$) is the lattice (magnon) specific heat per unit volume, $v_{ph}$ ($v_m$) is the sound velocity of acoustic phonons (group velocity of magnons) and $l_{ph}$ ($l_m$) is the phonon (magnon) mean free path as main unknown parameter. With decreasing temperature, the $l_{ph}$ ($l_m$) values gradually increase because of decreased populations of both quasiparticles (acoustic phonons and magnons) and, consequently, less collisions among them. In high-quality single crystals, the free path becomes finally limited by dimensions of the sample, based on Casimir theory that non-ideal (coarsened) crystal boundary absorbs heat of impacting quasiparticles and emits new equilibrium spectrum [22] (for experimental confirmation see Ref. 23). Considering the phonon term of thermal conductivity $\kappa_{ph}$, theoretical $\kappa_{ph} \sim T^3$ temperature dependence is obtained based on standard $T^3$ variation of lattice heat capacity, constant velocity of acoustic phonons and phonon-frequency independent mean free path given by Casimir. Similarly, the low-temperature magnon term is proposed (and indeed observed on CdCr$_2$Se$_4$) to be $\kappa_m \sim T^2$, due to $T^{3/2}$ variation of magnon specific heat and $T^{1/2}$-dependent effective velocity of magnons.

The decline of the CdCr$_2$Se$_4$ phonon thermal conductivity $\kappa_{ph} \sim T^{2.3}$ from the simple theoretical $T^3$ dependence (conditioned by temperature independent scattering) points to an existence of frequency-dependent scattering mechanisms influencing thermal conductivity down to very low temperatures. First, it can be noted that, even within the Casimir model, the deviation from $T^3$ can be obtained as shown already by Ziman who considered possibility of some specular reflection at sample boundaries of given roughness, leading to a frequency dependent ratio of the diffusive and specular reflection terms. As natural defects in crystals are concerned, a very specific behavior of phonon transport has been observed in metals and alloys namely due to dislocations that are supposed to move in quasi-free manner under striction strain of dynamic elastic wave represented by phonon (the movement called by Ziman as "fluttering" [24]). The energy dissipated is then proportional to the phonon frequency $\omega$ and if dominant, this mechanism naturally leads to the $T^2$ variation of phonon part of the metal thermal conductivity (see also Ref. 25). Recently, Kim et al., in their optimization of the Bi$_2$Te$_3$-type thermoelectric alloys, demonstrated a technology leading to a "mosaic crystal", where individual blocks are separated by low-angle boundaries with dense periodic arrays of dislocations. Here again, strong $\omega$-dependent phonon scattering is obtained but, instead of the freely moving dislocations, the mechanism relies on the static strain they impose [26,27].

Remarkable changes in phonon transport compared to pure crystalline systems were observed also by Pohl and his colleagues on numerous alkali halides with complex anionic substituents. For KBr:NO$_2$ single crystals in particular, the scattering mechanisms have been identified with rotation, liberation and tunnelling processes at NO$_2^-$ sites, which provide a broad spectrum of low-energy excited levels. Such local excitations are responsible for thermal conductivity strongly suppressed already from the phonon peak temperature with apparent $T^2$ dependence down to helium temperature but below $\sim$1 K, the thermal conductivity starts to recover [28]. The present polycrystalline sample of CdCr$_2$Se$_4$ shows low-temperature thermal conductivity very similar to KBr:NO$_2$, both in magnitude and temperature dependence. Scattering mechanisms, however, have now origin in grain boundaries and other associated imperfections, which complicate a clear understanding and simple modeling of thermal conductivity. On the other hand, any effort in this direction is highly desirable, since the phonon grain boundary scattering can suppress the transmittivity of long-wavelength phonons and, in the case of nanosize grains, the thermal conductivity can be largely reduced even at ambient temperatures. The nanostructuring is thus an effective tool for improved performance of thermoelectric materials [29,30].

Before presenting relevant experiments on CdCr$_2$Se$_4$, we consider as important to note some illustrative examples of the combined role of various volume defects associated with the grain boundaries and sample porosity which are projected to thermal transport of "real" polycrystalline ceramics.

(i) There are some cases when thermal conductivity is largely reduced in polycrystalline materials, but still follows the $T^3$ dependence at low temperatures, pointing to a constant (frequency-independent) mean free path. It is often presumed that the value obtained can be identified with grain size. While this is likely valid for magnon transport, considering that the magnon origin is mostly in the nearest-neighbor exchange interactions, the phonons can traverse the boundary depending on their wavelength (see below). The observation of ideal $T^3$ dependence need not to reflect the thermal resistance of the boundary itself but is rather related to reduced connectivity of grains due to sample porosity.

(ii) More attention has been given to cases when phonon thermal conductivity at low temperatures changed towards $\kappa_{ph} \sim T^2$ dependence. This is observed experimentally on many ceramic materials, including present CdCr$_2$Se$_4$. A special reference can be made to thorough research on nanocrystalline silicon of average grain sizes ranging from 550 to 64 nm by Wang et al. [31]. The authors have shown that $T^2$ trend observed at presence of important grain boundary scattering can be formally explained by a model in which transmittivity of phonons (or equivalently, the effective value of their mean free path) depends on phonon frequency as $\sim \omega^{-1}$. This means that only the phonons of low frequencies (long wavelengths) traverse through grain boundaries; see also the atomistic simulations of phonon transmittivity/reflectivity at grain boundaries of nanocrystalline Si and SiGe by Yang and Minnich [30].

(iii) In an attempt to get better physical insight into the scattering, the model of grain boundaries as dense array of dislocations has been applied by Snyder and his colleagues for fitting over broad temperature range of both the silicon data of Wang and own data on thermoelectrical alloys [27]. The authors have shown, with reference to earlier theoretical treatment of Klemens [32,33], that the low-temperature $T^2$ dependence can be explained as an effect of static strain imposed by dislocations on incoming phonons, leading to the $\omega$ dependent scattering term with magnitude given by actual density of dislocations in grain boundary.

(iv) As far as the studies of magnon thermal conductivity are concerned, the kinetic gas formula $\kappa_m = \frac{1}{3} C_m v_m l_m$ gives magnon contribution dependent on the energy averaged magnon velocity of $v_m$ and the magnon free path $l_m$. For rough estimate of $v_m$, at least for determination of its $T^{1/2}$-dependence, it is possible to consider the group velocity for thermal magnons at zero field ($E = \hbar \omega_k = k_B T$), given as $v_m(k_B T) = \frac{d\omega_k}{dk} = 2\sqrt{\frac{D\omega_k}{\hbar}} = \frac{2}{\hbar}\sqrt{Dk_B T}$, while rigorous derivation of the energy averaged velocity gives *1.62* times larger value, see detailed analysis by Pan et al. [34]. In a model of temperature independent $l_m$, the magnon thermal conductivity should thus follow quadratic $T^2$ temperature dependence as mentioned for the first time by McColum et al. [35]. An open question remains the transmittivity of magnons through grain boundaries. The study of various forms of YIG by Miura et al. [36] has shown that thermal conductivity of single-crystalline sample at 2 K could be at least 55% ascribed to the magnon contribution. This ratio has been gradually diminished for the polycrystalline sample with 20 μm grains to ~30%, and for samples with 2 and 0.5 μm grains, the magnon contribution to macroscopic heat transport was below detection limit. This suggests that there is no or only a very limited magnon penetration through grain boundaries, and, therefore, in ceramic samples, the heat is predominantly transmitted by the long-wavelength or surface phonons.

**Low-temperature thermal conductivity of densified $CdCr_2Se_4$ sample**

The original material, whose properties we presented in Fig. 4, was prepared with relatively large porosity of 17%. To sort out effects of the porosity and the grain boundary itself on the thermal conductivity, the material was subjected to a densification by spark plasma sintering at 300 MPa, followed with annealing at 500 °C in selenium atmosphere for 24 hours. This resulted in the sample porosity largely reduced to 5%. As seen in upper and lower panels of Fig. 5, the dense sample shows thermal conductivity with significantly enhanced phonon peak and four-times larger magnitude when compared to the porous sample. A steep drop is observed with increasing temperature, ending by an anomaly at $T_C$ ~130K. No tendency to the $T^{-1}$ trend can be identified. In our understanding, the reason for such behaviors is in the coupling of lattice dynamics with ferromagnetic order in $CdCr_2Se_4$, and the observed drop in the 70-130 K range reflects a gradual increase of ferromagnetic fluctuations. Above Curie temperature, the thermal conductivity follows the $T^{-1/2}$ trend, similar to our porous sample but with twice larger magnitude. Such seemingly analytical dependence is likely only accidental and specific for the present system in paramagnetic state. It may result from a combination of the phonon-phonon Umklapp processes, atomic displacements due to local spin correlations and some defects associated with the microstructure of our polycrystalline sample. In this context, it is worth mentioning the first-order magnetostructural transition in CrN which represents an extreme case of the coupled lattice dynamics and spin arrangement. That is clearly manifested in the thermal conductivity showing incomparably large drop when approaching the antiferromagnetic Néel temperature, followed with practically constant or even slightly increasing trend at higher temperatures, see e. g. Fig.6 in Ref.17.

Behaviors of the dense and porous samples below the phonon peak are illustrated in more details in the lower panel of Fig. 5. One may note that, although the low-temperature thermal conductivity of dense sample has increased four times, its suppression in field of 13 T remained similar, pointing to practically unchanged ratio of the magnon term to the phonon one. These results suggest that porosity influences thermal conductivity, namely at magnetically polarized state just as a multiplying factor. Phenomenological models anticipate a reduction with increasing porosity *p* either exponential one or as *(1-p)^t* when percolation model is implemented [37]. More physical approach is to implement the effective medium theory, devised originally for treatment of electric conduction and galvanomagnetic phenomena in inhomogeneous materials, see e. g. [38]. For the simplest model of a two-phase composite with random distribution of spherical crystallites the solution has been presented already in 1950-ies and tested successfully on experimental data for various alloy mixtures [39]. Analogical model has been recently used for both the electric and thermal conductivity of hot-pressed $(Bi,Sb)_2Te_3$ thermoelectrics, considering their structure as a mixture of conducting material and insulating pores [40]. Considering all these scenarios, we may conclude that at least in first approximation, the conductivity is reduced with porosity linearly for small *p*. This allows us to estimate, at absence of any real single-crystal specimen, that a hypothetical, porosity-free $CdCr_2Se_4$ sample, would show thermal conductivity at least five-times larger than our original porous sample (*p*=0.17) and ~20% larger compared to the densified one (*p*=0.05).

Taking this fact into account, we analyze first the magnon term $\kappa_m$ of thermal conductivity. The lower panel of Fig. 5 enables to evaluate the difference between measured $\kappa_{0T}$ and $\kappa_{13T}$ (including the porosity correction) as $\kappa_m = 0.055$ Wm$^{-1}$K$^{-1}$ at 2 K. At the same time, we note that the magnon heat capacity $C_m= 0.0537$ Jmol$^{-1}$K$^{-1}$ ($7.42\cdot10^4$ Jm$^{-3}$K$^{-1}$), velocity of dominant magnons in for CdCr$_2$Se$_4$ with spin-wave stiffness of $D = 33.5$ meVÅ$^2$ given by $\langle v_m \rangle = 1.62 \frac{2}{\hbar}\sqrt{Dk_BT} = 840\sqrt{T} = 1190$ ms$^{-1}$ and actual grain size of ~1 μm give an estimate of grain-boundary limited $\kappa_m = \frac{1}{3}C_m v_m l_m = 0.221$ Wm$^{-1}$K$^{-1}$ at 2K. This four-times larger value compared to the observed one would mean that the effective magnon path is apparently only one quarter of the grain size, i. e. about 0.25 μm. This corroborates the assumption of no penetration of magnons through grain boundaries. In a more realistic explanation, the present result should be related rather to a large refraction power of grain boundaries, so that the spin waves are mostly localized in grains and only one quarter of their incident heat flux is transmitted further. It is of interest that despite the complexity of heat transfer in polycrystalline CdCr$_2$Se$_4$, the measured magnon contribution follows the ideal $\kappa_m \sim T^2$ dependence (at least in the lowest $T$=2-6 K range, see Fig. 4b). This points to a frequency independent reflection coefficient for incident long-wavelength magnons, the value of which ($R$=75%) is much larger that commonly reported for phonon reflection at grain boundaries. In analogy to backscattering of acoustic waves at the two-media interface as applied in the theory of grain-boundary thermal resistance [41], we are considering that the inter-grain region can be viewed as defective and less compact medium of significantly reduced spin-wave stiffness $D'$ and the grain-boundary refraction is predominantly determined by relative change of the magnon group velocity, $R \sim \left(\frac{v_m - v_{m'}}{v_m}\right)^2 = \frac{D-D'}{D}$.

The phonon part of thermal conductivity, evidenced in the lower panel of Fig. 5 as $\kappa_{13T}$, makes after porosity correction a value of $\kappa_{ph} = 0.110$ Wm$^{-1}$K$^{-1}$ at 2 K. Applying the kinetic gas formula $\kappa_{ph} = \frac{1}{3}C_{ph}v_{ph}l_{ph}$ with heat capacity $C_m= 0.0072$ Jmol$^{-1}$K$^{-1}$ ($1.00\cdot10^4$ Jm$^{-3}$K$^{-1}$) and sound velocity v=2400 m/s, we estimate the phonon mean free path to be effectively ~1.4 μm which corroborates the estimated average grain size. The large difference in mean free path between magnons and phonons suggests that magnons are much more sensitive to "structural imperfections" than phonons. This is rather unexpected result since based on the intrinsic scattering mechanisms and experiments on single-crystal materials, the magnon free path generally exceeds that of phonons.

Very short magnon free path we deduced above for porosity-free CdCr$_2$Se$_4$ polycrystalline material (effective values $l_m$= 0.25 μm, $l_{ph}$= 1.4 μm at 2 K) can be understood by considering that apart from magnetic defects (e.g. the presence of Cr$^{2+}$ due to selenium vacancy V$_{Se}$) there are random atomic displacements associated not only with grain boundaries, but with coalescing crystallites, dislocations etc. Even if small, such atomic displacements may influence dramatically magnetic interactions, specifically in systems where competing exchange paths exist between neighboring spins, the strengths of which depend on different hopping integrals when tight-binding model is used [42]. As an illustrative example how orbital overlaps and, consequently, the hopping integrals vary with interatomic distances, we refer to recent study of magnetostructural transition in the CrN system possessing rock-salt structure [3]. Similarly, the present CdCr$_2$Se$_4$ system is subjected to the competition between the antiferromagnetic direct Cr-Cr exchange (dependent on interatomic distance as $d_{Cr-Cr}^{-10}$) and the 90° ferromagnetic Cr-Se-Cr superexchange (dependent as $d_{Cr-Se}^{-14}$). Consequently, the dominance of ferromagnetic superexchange can locally switch to antiferromagnetic exchange, which results in a strong magnon scattering.

Since phonons are shown to dominate the inter-grain transport of $CdCr_2Se_4$, the observation of the porosity-independent magnon/phonon ratio of bulk thermal conductivity, see Fig. 5(b) for dense and porous sample, is no longer surprising. This means that the reduction with porosity need not be result of a mere geometric factor (effective cross-section), but there can be other scattering mechanism, frequency dependent or independent, acting on traversing phonons. Here, in analogy to Ziman's "fluttering" of quasi-free dislocations in single crystals, we anticipate that surface atoms of $CdCr_2Se_4$ crystallites/grains can occupy multiple sites with certain barriers between them, and under elastic strain of incoming phonons they are "rolling around" as pebbles at windy shore. However, we cannot estimate, before proper quantum-mechanical treatment is available, how large the scattering effect on phonons can be. Another possible mechanism to consider also in porosity-free polycrystals is the dislocation strain described by Kim et al. [27]. Its strength is, however, hardly predictable for common samples that possess high-angle grain boundaries.

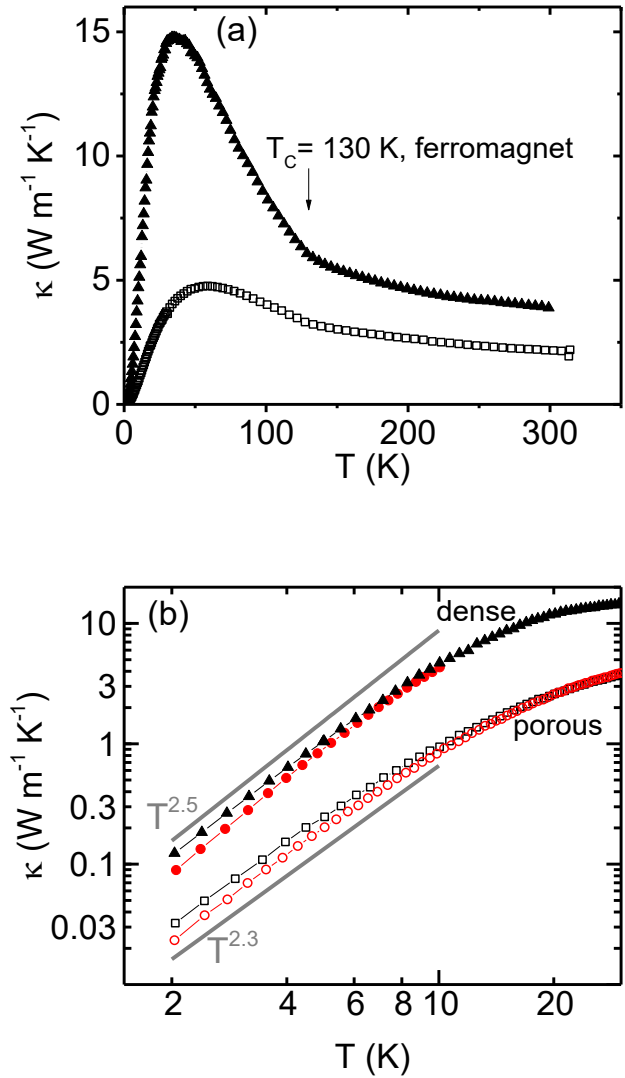

Fig. 5. (a) Thermal conductivity on dense $CdCr_2Se_4$ sample (5% porosity, solid triangles) compared to the original porous one (17% porosity, open squares). (b) The effect of external field of 13 T at low temperatures (red symbols).

Finally, we are returning to the suppression of thermal conductivity in magnetic field and the $T^2$ dependence of magnon contribution obtained for present samples. First, there is a question whether the application of magnetic field influences also the phonon contribution of ferromagnetic CdCr$_2$Se$_4$, e. g. due to magnetization of volume impurities containing paramagnetic ions, which might bring additional stress to the main phase. Considering that amount of foreign phases in present samples have been found to be negligible (in maximum ~1%), we argue that at least at lowest temperatures, the data obtained at 13 T correspond well to $\kappa_{ph}$ at zero field. With increasing temperature to ~8 K, the situation becomes more complicated. As illustrated in the inset of Fig. 4a, the $\kappa_{0T}$ - $\kappa_{13T}$ data, even after correction for incomplete reduction of magnons, cease to follow the expected $T^2$ trend of $\kappa_m$. There is also some change in the character of isothermal $\kappa_m(H, T \geq 10\ K)$ curves, where instead of the theoretically-predicted gradual quenching exemplified in Fig. 4b for 2.5 K and scaled as $H/T$, certain drop of magnon thermal conductivity has been observed in low fields, followed with nearly constant value in 1 – 13 T [7]. First, let us note that the $T^{2.3}$ dependence of phonon thermal conductivity suggests the $l_{ph}$ for our porous sample to be effectively reduced with temperature as $T^{-0.7}$. This predicts its decrease from 1.4 µm at 2 K to ~0.5 µm at 8 K while for dense sample ($T^{2.5}$ dependence) a slower decrease is anticipated but still below the grain size. At the same temperature (~8 K), the velocity of dominant magnons reaches that of acoustic phonons (sound velocity of 2400 m/s). One may thus expect the onset of some phonon-magnon scattering and in our opinion, the applied magnetic field brings two opposing effects which can explain the peculiar behavior of $\kappa_{0T}$ - $\kappa_{13T}$ above 8 K. In particular, we consider that the predicted suppression of magnon conductivity due to external magnetic field can be partially compensated by an increase of lattice contribution due to decreased phonon scattering on magnons. With increasing temperatures above ~20 K, we expect that the effect of phonon-magnon scattering becomes gradually unimportant since the magnon population and their contribution to thermal conductivity though increasing, are overwhelmed by a still steeper increase and dominance of phonon contribution.

**CONCLUSION**

The present study was focused to the thermal conductivity in the ferromagnetic thiospinel CdCr$_2$Se$_4$. Two polycrystalline samples with different porosities were investigated – the first one prepared by standard sintering (porosity 17%) and second one densified by spark plasma procedure (porosity 5%). The thermal conductivity experiments were performed in the range from room temperature down to 2 K. To distinguish the role of phonons (standard lattice contribution) from that of magnons, additional measurements in external magnetic fields up to 13 T were done below 20 K. The results were discussed with reference to older and recent literature sources describing numerous scattering mechanisms that refer to both the intrinsic (phonon-phonon, phonon-magnon) processes and defect/microstructure related extrinsic ones.

As a noteworthy and rather unexpected result, it can be mentioned that the mean free path for magnons at the lowest measured temperatures is well below the CdCr$_2$Se$_4$ grain size. Such observation is related to the presence of local spin fluctuations arising due to small variations of interatomic distances. The reason for such strong coupling between the spin and lattice degrees of freedom is in a competition of the antiferromagnetic Cr-Cr direct exchange and the 90° ferromagnetic Cr-Se-Cr superexchange, each of specific dependence on local atomic displacements. On the other hand, the mean free path of phonons is about four-times larger and seems to be limited by grain-boundary scattering and due to its slight decrease with increasing temperature, the contribution to the low-temperature thermal conductivity declines from the ideal $T^3$ trend to $\kappa_{ph} \sim T^{2.5}$ for the dense sample and $\sim T^{2.3}$ for the porous one.

Another specific feature is behavior of thermal conductivity above the phonon peak at ~50 K. Instead of commonly observed $T^{-1}$ decrease there is much steeper drop ending with an anomaly at ferromagnetic $T_C$ ~130 K. With further temperature increase up to ~300 K, the thermal conductivity decreases only slightly, following an apparent $T^{-0.5}$ trend. The behaviors mentioned point again to a strong coupling between the lattice dynamics and spin states. It is obvious that the spin fluctuations at approaching $T_C$ and local spin correlations surviving above $T_C$ influence significantly the temperature dependence of thermal conductivity in $CdCr_2Se_4$.

**ACKNOWLEGEMENTS**

We acknowledge the Operational Program Research, Development and Education financed by the European Structural and Investment Funds and by the Ministry of Education, Youth and Sports (MEYS) of the Czech Republic, Grant No. CZ.02.01.01/00/22 008/0004594 (TERAFIT). The experiments were performed at Materials Growth & Measurement Laboratory (MGML) supported within the program of Czech Research Infrastructures (project no. LM2018096). Computational resources were provided by the MEYS of the Czech Republic through the e-INFRA CZ (ID:90254). S. A. B. and E.G. thank the Normandy Region for financial support.

**ASSOCIATED CONTENT**

**Supporting Information**
Supporting Information is available at
< link to Supporting Information will be inserted>
Additional Experimental information, namely concerning the sample properties (XRD, microstructure…) is provided.

**Data availability**
Data associated with this study are available from Zenodo at doi.org/10.5281/zenodo.18622579.

TOC picture

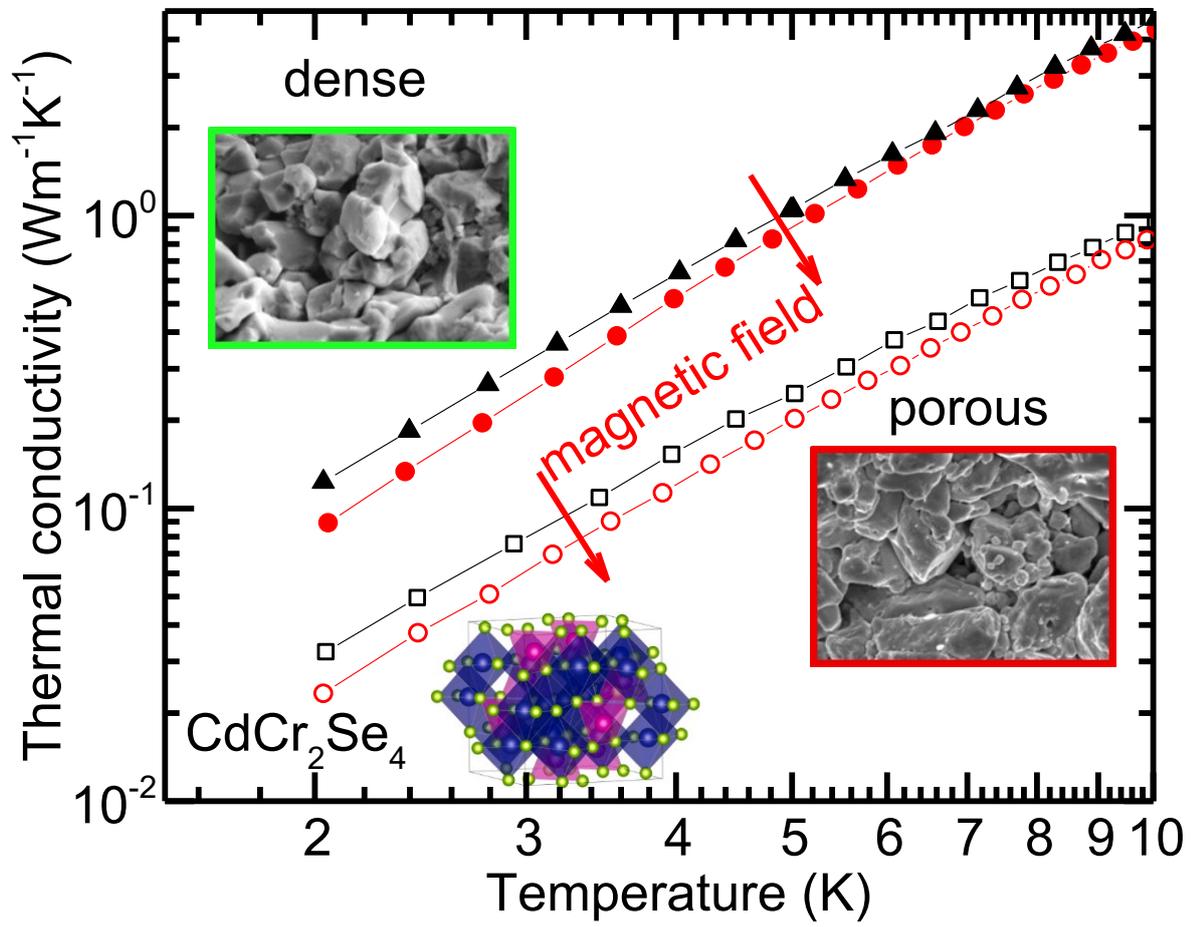

# Supplementary information

# Thermal conductivity of $CdCr_2Se_4$ ferromagnet at low temperatures: role of grain boundaries and porosity

Jiří Hejtmánek[1*], Kyo-Hoon Ahn[1], Zdeněk Jirák[1], Petr Levinský[1], Jiří Navrátil[1], Sandy Al Bacha[2], Emmanuel Guilmeau[2] and Karel Knížek[1]

[1]Institute of Physics of the Czech Academy of Sciences, 162 00 Prague 6, Czech Republic.
*Corresponding author:* hejtman@fzu.cz

[2]CRISMAT, CNRS, Normandie Univ, ENSICAEN, UNICAEN, 14000 Caen, France

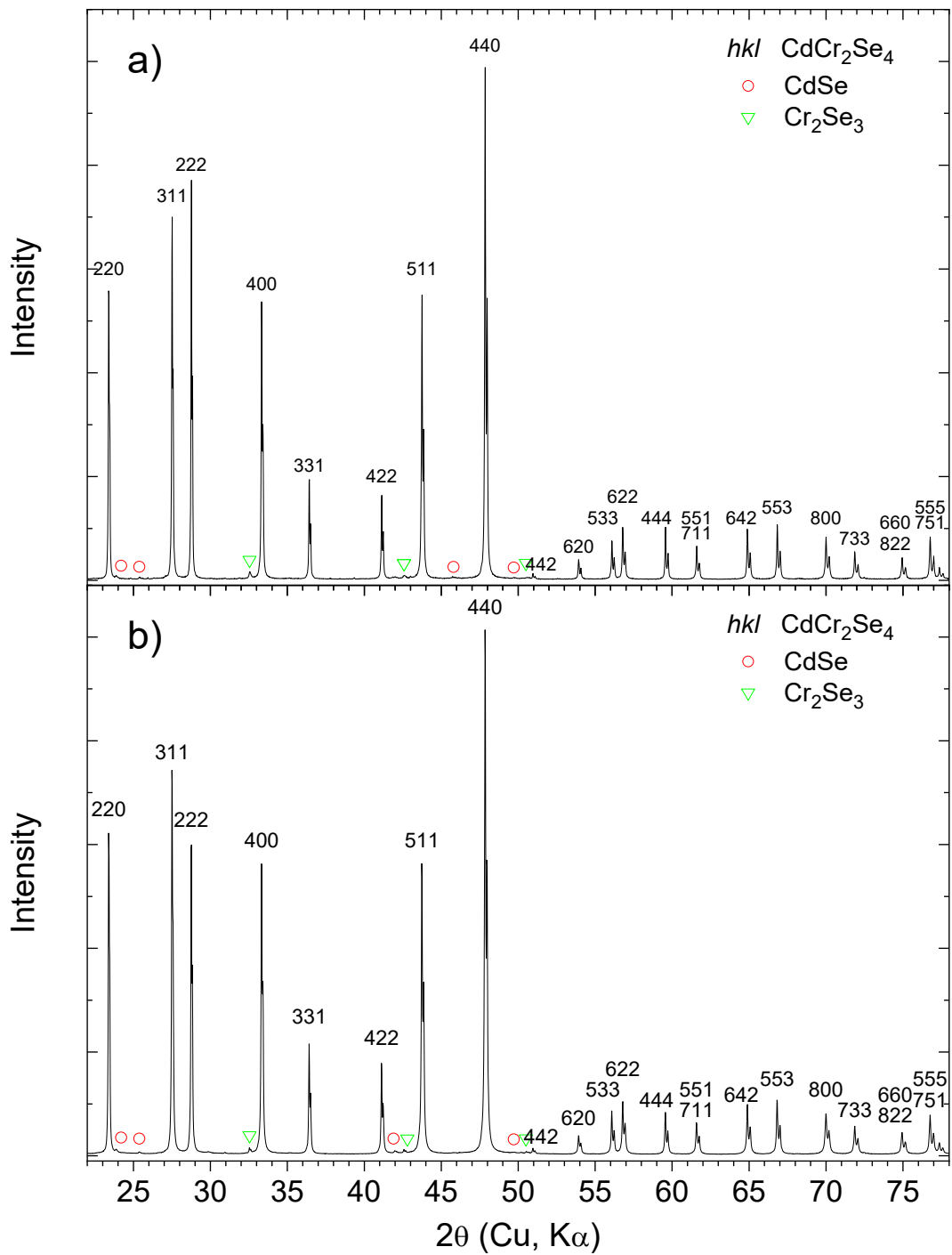

Fig. S1. Powder X-ray diffraction patterns of CdCr$_2$Se$_4$ samples, a) before SPS sintering, b) after SPS sintering and selenization (i.e. the "dense" sample). The *hkl* indexes of the main reflections and main peaks of impurity phases are indicated. The volume amount of each minority phase, CdSe (o) and Cr$_2$Se$_3$ (∇), is less than 1%.

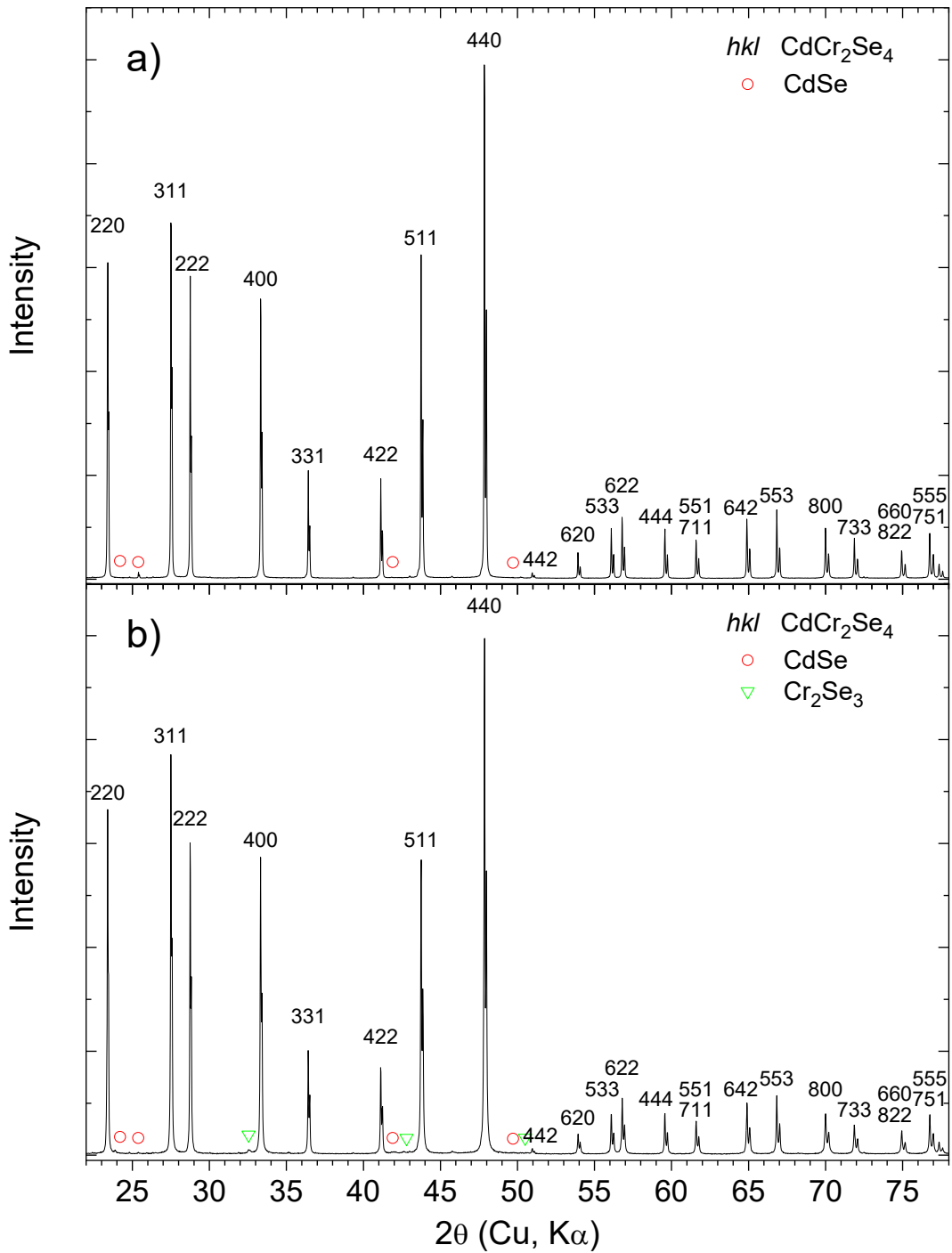

Fig. S2. Powder X-ray diffraction pattern of the hot-pressed (HP) "porous" CdCr$_2$Se$_4$ sample. a) powder sample before HP sintering. b) solid sample after HP sintering. The *hkl* indexes of the main reflections and main peaks of impurity phases are indicated. The volume amount of each minority phase, CdSe (o) and Cr$_2$Se$_3$ (∇), is less than 1%.

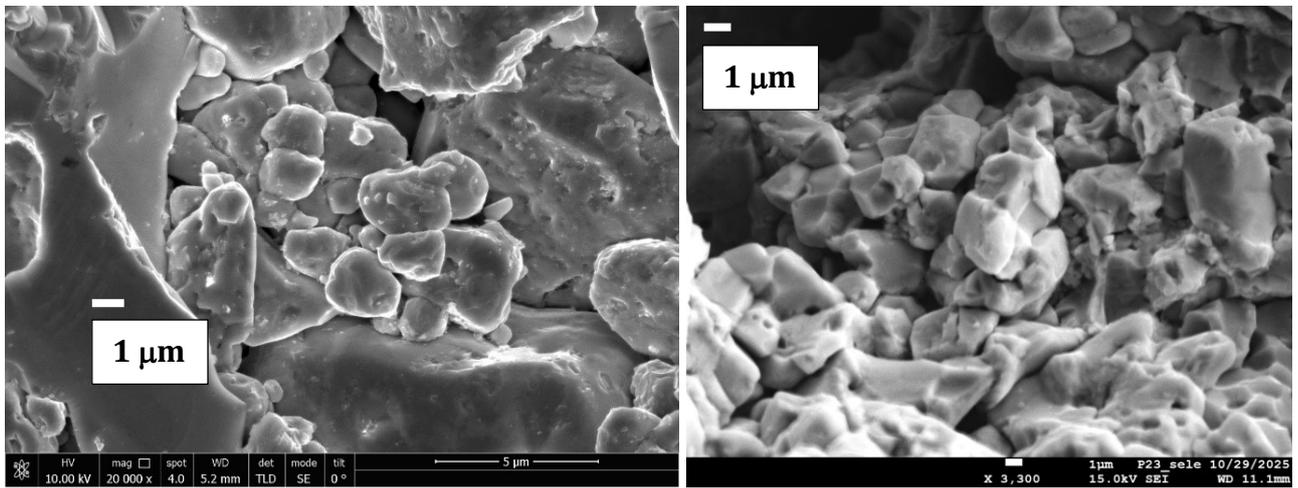

Fig. S3. Scanning electron microscope images (secondary electrons) of the samples. Left: hot-pressed porous sample (83% density), right: SPS sintered ceramics with post annealing in selenium vapor, i.e. the dense sample (95% density).

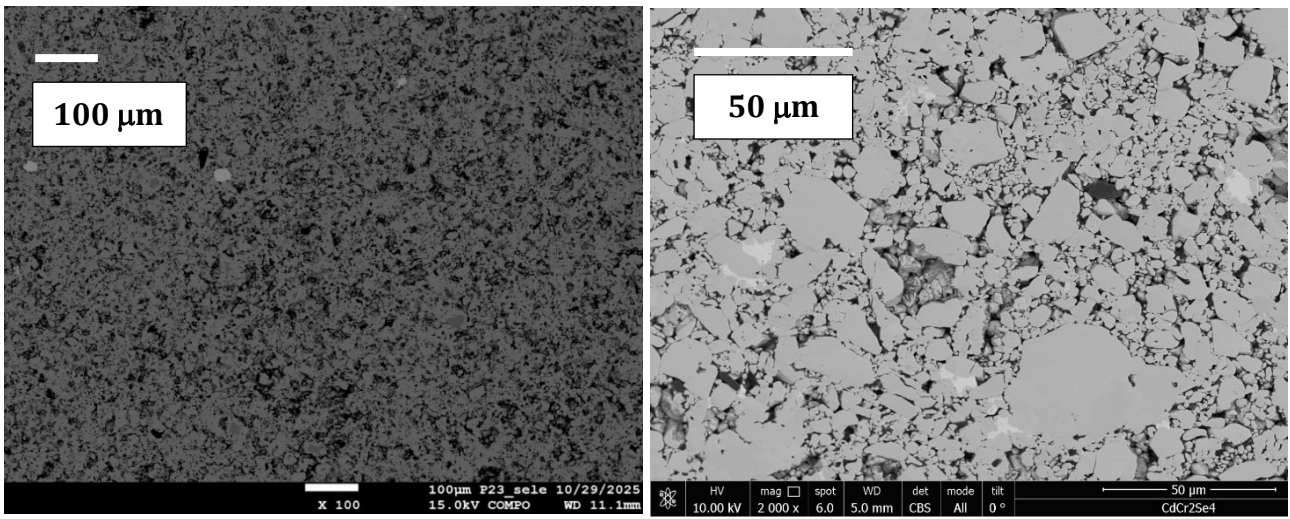

Fig. S4. Scanning electron microscope images (backscattered electrons from polished surface). Left: hot-pressed porous sample (83% density), right: SPS sintered ceramics with post annealing in selenium vapor, i.e. the dense sample (95% density).

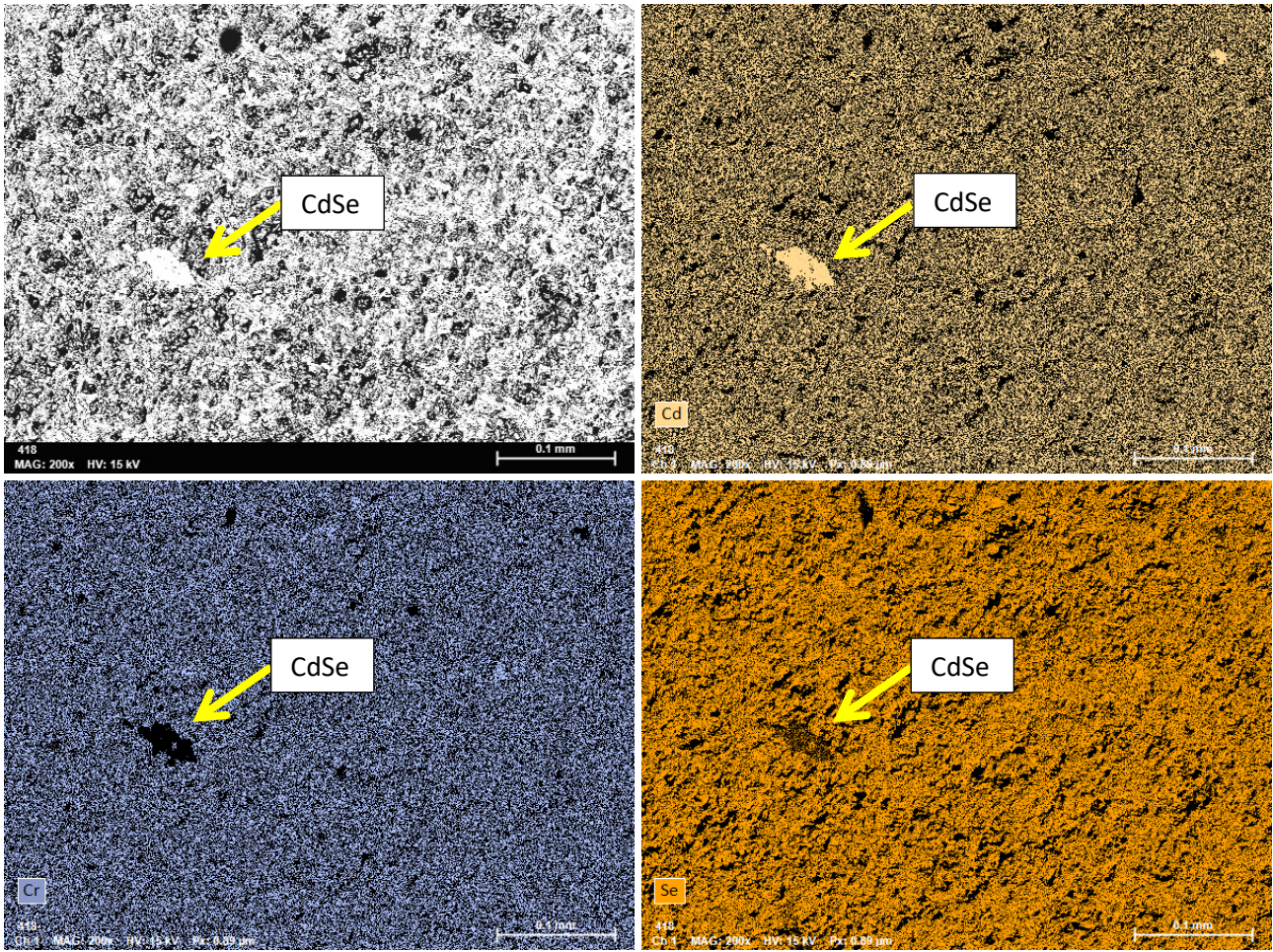

Fig. S5. Scanning electron microscope image and elemental X-ray maps of the selenized SPS sintered "dense" sample. The morphology of CdSe impurities, represented by large, isolated grains, guarantees their negligible role in thermal conductivity which is of "percolative" character. The $Cr_2Se_3$ impurities are represented by similar, spatially isolated grains with the size of units of micrometers.

Table S1. Energy-dispersive X-ray analysis of several points on the dense sample. The data was collected using the Jeol JXA-8230 instrument equipped with a Bruker QUANTAX 200 spectrometer. Quantitative analysis was performed with the method PB-ZAF.

| Atomic % | | | | | Normalized to 7 atoms | | | |
|---|---|---|---|---|---|---|---|---|
| Cd | Cr | Se | sum | | Cd | Cr | Se | |
| 14.50 | 28.06 | 57.44 | 100.00 | | 1.02 | 1.96 | 4.02 | |
| 14.80 | 28.00 | 57.20 | 100.00 | | 1.04 | 1.96 | 4.00 | |
| 14.90 | 28.38 | 56.73 | 100.01 | | 1.04 | 1.99 | 3.97 | |
| 14.01 | 29.90 | 56.09 | 100.00 | | 0.98 | 2.09 | 3.93 | |
| 14.12 | 29.83 | 56.05 | 100.00 | | 0.99 | 2.09 | 3.92 | |
| | | | | | **1.01** | **2.02** | **3.97** | mean |
| | | | | | 0.02 | 0.06 | 0.04 | ± |